\RequirePackage[T1]{fontenc}
\documentclass[conference]{IEEEtran}
\IEEEoverridecommandlockouts
\usepackage{cite}
\usepackage{amsmath,amssymb}
\usepackage{graphicx}
\usepackage{booktabs}
\usepackage{url}
\usepackage[hidelinks]{hyperref}
\usepackage{textcomp}
\usepackage{xcolor}
\def\BibTeX{{\rm B\kern-.05em{\sc i\kern-.025em b}\kern-.08em T\kern-.1667em\lower.7ex\hbox{E}\kern-.125emX}}

\begin{document}

\title{When Should a Wearable Camera Record?\\
EEG-Guided Context Capture Driven by a Sustained Brain State}

\author{%
\IEEEauthorblockN{Namai Chandra}
\IEEEauthorblockA{\textit{Indian Institute of Technology Madras}\\
Chennai, India\\
23f3000200@es.study.iitm.ac.in}
\and
\IEEEauthorblockN{Dong Anh Khoa To}
\IEEEauthorblockA{\textit{VinUniversity}\\
Hanoi, Vietnam\\
22khoa.tda@vinuni.edu.vn}
\thanks{\copyright~2026 IEEE. Personal use of this material is permitted. Permission from IEEE must be obtained for all other uses, in any current or future media, including reprinting/republishing this material for advertising or promotional purposes, creating new collective works, for resale or redistribution to servers or lists, or reuse of any copyrighted component of this work in other works.}
}

\maketitle

\begin{abstract}
Selective recording can reduce the amount of wearable-camera footage a user must store and review. We use a sustained task-versus-rest state estimated from electroencephalography as a recording trigger and measure how smoothing the score changes which footage the camera would retain. In offline replay, a per-person band-power model produces a continuous score, and a three-second persistence rule requests fixed ten-second clips. Across 45 held-out working-memory blocks from fifteen sessions and five people, exponential smoothing with a one-second half-life raised the probability of ranking a task window above a rest window from 76.6\% to 79.8\%. It reduced requested footage by 23\% and retained at least five seconds of every task block. Total retained task time fell from 902 to 697 seconds. In 51 separate surprise-task recordings, an eight-electrode scalp-layout proxy achieved 82\% ranking performance, compared with 84\% for the full montage and 61--65\% for eyeglass positions. The montage comparison informs electrode placement for surprise detection. The replay shows why recording policies should be evaluated in seconds of footage as well as by score discrimination.
\end{abstract}

\begin{IEEEkeywords}
electroencephalography, wearable camera, lifelogging, passive brain--computer interface, event-triggered recording
\end{IEEEkeywords}

\section{Introduction}
\label{sec:intro}

Lifelogging devices such as SenseCam showed that a passively captured visual record can help people revisit their day~\cite{sensecam,sellen}. Continuous video consumes battery and storage and increases review effort. Manual capture requires the wearer to stop and act, while automatic capture requires a rule for when the camera should turn on.

\begin{figure*}[t]
\centering
\includegraphics[width=\textwidth]{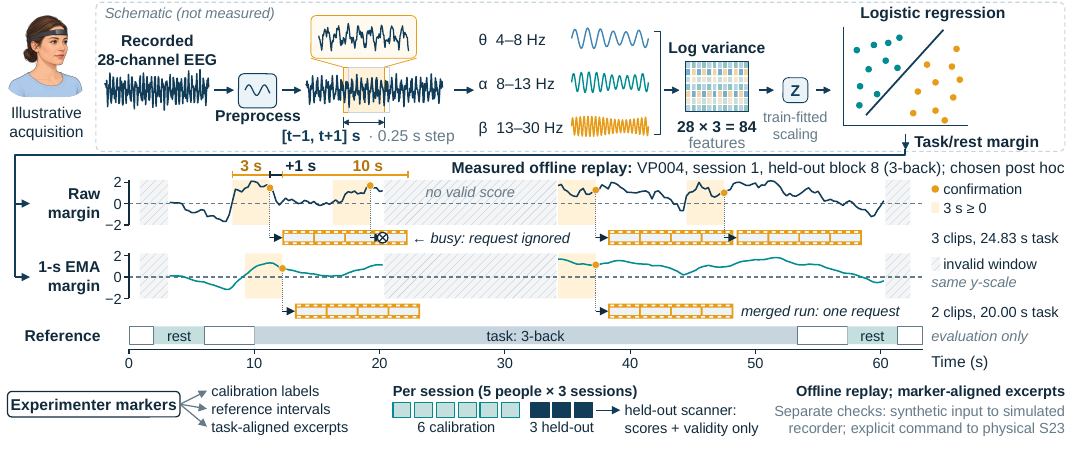}
\caption{One real held-out block (VP004, session~1, block~8). (a)~From scalp to score. (b)~Raw score (navy) and 1~s smooth (teal) on the same axis; bars are the 10~s clips each version would record.}
\label{fig:pipeline}
\end{figure*}

We study whether a sustained task-versus-rest state estimated from electroencephalography (EEG) can drive recording requests (Fig.~\ref{fig:pipeline}). We replay raw and smoothed scores from the same fitted model through the same recording rule to measure how smoothing affects the footage requested and the task time retained. A separate surprise-task analysis examines electrode count and placement.
\begin{itemize}
\item An offline EEG-to-recording-request pipeline: 84 band-power features and a per-person classifier give a task-versus-rest score, and a persistence rule (score elevated for three seconds) turns it into ten-second recording requests.
\item A paired evaluation of the recording decision on 45 held-out task blocks from five people: exponential smoothing with a one-second half-life lowers recording demand by 23\% overall, reduces demand in every participant, and preserves at least five seconds of every task block, while retaining about one fifth less task footage.
\item An electrode comparison on 51 surprise-task recordings, reported as ranking probabilities: eight electrodes in a headband-layout proxy keep 82\% against 84\% for the full montage, while eyeglass positions fall to 61--65\%. For this task, placement matters more than count; the workload montage remains untested.
\item Negative results that shaped the design: the tested brief-event detectors did not exceed a random-placement null, and a decoder that separates stimulus-locked epochs failed on a continuous stream.
\end{itemize}

\section{Related Work}
\label{sec:related}

\textbf{Deciding when a lifelogging camera records.} SenseCam used light, motion and periodic triggers for still capture~\cite{sensecam}, and controlled studies found benefits for recall~\cite{sellen}. Bruun and Stentoft studied participants' experiences of wearable-camera lifelogging~\cite{bruun}. HeartbeatCam uses a smartwatch stress signal to trigger a camera~\cite{heartbeat}, and StartleCam used skin conductance to preserve images~\cite{startlecam}.

Memento uses EEG to infer emotions and trigger lifelogging~\cite{memento}. We use a sustained task-versus-rest score to request fixed-length clips. In offline replay, we hold the classifier and recording rule fixed and vary only score smoothing. Across 45 held-out blocks, smoothing reduced requested footage from 980 to 750 seconds. It retained at least five seconds of every task block, while total retained task footage fell from 902 to 697 seconds.

\textbf{Passive brain--computer interfaces and workload.} Passive BCIs read ongoing EEG without asking the user to issue commands~\cite{zander}. Mental workload is one of their better established targets: frontal theta power rises and parietal alpha power falls when a person engages in a demanding task, and these changes are consistent enough to transfer across tasks within a person~\cite{ke,gallegos}. Band power remains a competitive baseline, which is why we adopt it. The passive-BCI literature, however, reports how well a classifier separates two states. A camera requires a sequence of start and stop decisions derived from the classifier score, and the value of those decisions depends on how a noisy score is converted into actions, a point made for asynchronous BCIs by utility metrics that charge for every false activation~\cite{utility}.

Calibration and implementation cost are separate design questions. Reviews of calibration reduction describe the effect of participant and session variability on EEG decoding~\cite{calibration}. EEGNet learns compact convolutional representations~\cite{eegnet}, and Q-EEGNet addresses quantized execution on low-power hardware~\cite{qeegnet}. Our linear model isolates the recording-rule comparison; it is not a measured accuracy or energy comparison against those networks. Reproducible benchmarking~\cite{moabb} and small-sample cross-validation uncertainty~\cite{varoquaux} also motivate keeping task/rest replay, within-recording montage tests, and cross-person transfer results separate.

\textbf{Electrode budget.} Consumer devices carry between two and sixteen dry electrodes, and the positions a glasses frame can reach (brow and temples) are not the positions where workload signals are strongest. Which positions a wearable needs, and how a calibrated model degrades as electrodes are removed, has rarely been reported for a capture application. We measure it on a separate 16-channel surprise-task dataset, as a placement proxy rather than a workload result.

\section{Methodology}
\label{sec:method}

\subsection{Design principle: sustained states, not spikes}
Before building the recorder we asked which kind of brain event a few dry electrodes can detect reliably. Brief evoked responses are attractive, because a surprise or an insight is exactly the moment one would want to keep. We tested five brief-event detectors, from xDAWN spatial filtering to a topological descriptor, on four participants against a null in which the same number of flags were placed at random. None beat the null; the best, which reached twice its random baseline (six matched events against a random expectation of three), had $p=0.13$. Sustained task engagement, by contrast, was recoverable, and it is also the state in which a person is least willing to reach for a button. The system therefore reacts to states that persist for seconds. Event-related components such as P3a and P3b are characterised relative to stimulus onset~\cite{polich}; detecting their occurrence in an unmarked stream is a separate requirement from distinguishing labelled epochs.

\subsection{From scalp to score}
Signals are resampled to 200~Hz. A ridge regression from the two electro-oculography (EOG) channels, fitted on a pre-task segment and then frozen, subtracts the component predicted by the ocular channels~\cite{croft}; artifact windows (excursions above 120~$\mu$V, steps above 60~$\mu$V per sample, or a flat trace) produce no score. Butterworth filters isolate theta (4--8~Hz), alpha (8--13~Hz), and beta (13--30~Hz), and in a centred two-second window advanced every 0.25~s each channel contributes the log variance of each band, giving $28\times3=84$ features (Fig.~\ref{fig:smoothing}a,b). For channel $k$ and band $b$, the feature is
\begin{equation}
\phi_{k,b}(t)=\log\!\left(\max\{\operatorname{Var}_{u\in W_t}x_{k,b}(u),\epsilon\}\right),
\label{eq:features}
\end{equation}
where $x_{k,b}$ is the band-filtered signal, $W_t$ is the centred two-second window, and $\epsilon=10^{-12}$ is a numerical variance floor. Calibration-fitted standardization and class-balanced, L2-regularised logistic regression produce the signed margin
\begin{equation}
\boldsymbol z_t=\frac{\boldsymbol\phi_t-\boldsymbol\mu}{\boldsymbol\sigma},
\qquad m_t=\boldsymbol w^\top\boldsymbol z_t+b_0,
\label{eq:margin}
\end{equation}
with componentwise division and $\boldsymbol\phi_t\in\mathbb R^{84}$. The calibration means $\boldsymbol\mu$, standard deviations $\boldsymbol\sigma$, weights $\boldsymbol w$, and intercept $b_0$ are frozen before evaluation; positive $m_t$ means ``task-like''. Regularisation is chosen on two calibration folds from $\{0.01, 0.1, 1, 10\}$. On the same recordings and folds, Riemannian tangent-space features, dense Fourier bins, volatility features, a change-point model, and two Bayesian state-space smoothers all lost or tied in paired comparisons. The only addition that survived is the single smoothing parameter described next.

\paragraph{Personal calibration}
For $N$ valid calibration windows, let $y_r\in\{-1,+1\}$ denote rest or task and $N_y$ the number in class $y$. With class weight $\omega_y=N/(2N_y)$, the fitted linear model minimises
\begin{equation}
\min_{\boldsymbol w,b_0}\;
\frac{1}{2}\|\boldsymbol w\|_2^2+
C\sum_{r=1}^{N}\omega_{y_r}\log\!\left(1+e^{-y_r m_r}\right).
\label{eq:calibration}
\end{equation}
Here $m_r=\boldsymbol w^\top\boldsymbol z_r+b_0$. The weights balance task and rest contributions, while $C$ controls regularisation. Scaling and model selection use calibration blocks only; neighbouring windows are not randomly mixed across the calibration/evaluation boundary. The subsequent recording comparison reuses the same fitted margin stream, so a change in retained footage is attributable to the temporal rule rather than a newly trained decoder.

\subsection{From score to recording}
\begin{figure}[t]
\centering
\includegraphics[width=\columnwidth]{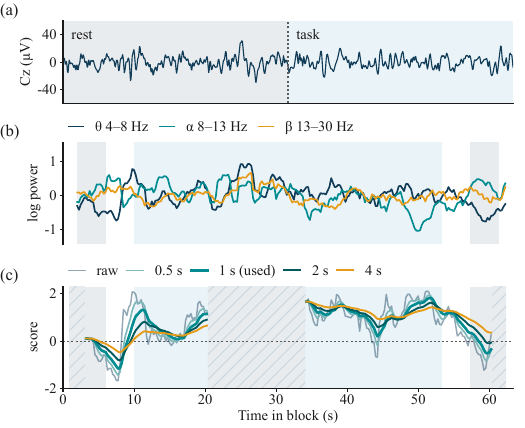}
\caption{The same real block as Fig.~\ref{fig:pipeline}. (a)~Recorded Cz around task onset. (b)~Band power, the model's inputs. (c)~Raw score and exponential smoothing at half-lives 0.5, 1, 2, and 4~s. The recorder uses the 1~s trace.}
\label{fig:smoothing}
\end{figure}

The recorder sees only the score stream and its validity. A fixed exponential moving average (EMA) with half-life $h$ smooths the margin,
\begin{equation}
s_t=(1-\alpha)\,s_{t-1}+\alpha\, m_t,\qquad \alpha = 1-2^{-0.25/h},
\label{eq:ema}
\end{equation}
with the history reset after a missing sample; $h=0$ denotes the raw score. Fig.~\ref{fig:smoothing}(c) shows the raw score and the 1~s smooth used in the recorder. When $s_t\ge0$ for $c$ consecutive seconds the state is \emph{confirmed}. One second later, accounting for the end of the centred feature window, the recorder requests one clip of fixed duration $L=10$~s. Zero-phase filtering still uses future samples, so this evaluation is offline replay. Requests arriving while a clip is active are ignored, and each confirmed interval may request at most one clip. In the main comparison $c=3$~s and only $h$ changes, from 0 to 1~s, with classifier weights, scaling, and validity masks held fixed (Fig.~\ref{fig:pipeline}b).

\paragraph{Clip acceptance}
Let $\tau_j$ be the first score time at which valid positive run $j$ has persisted for $c$ seconds. Its candidate clip starts at $a_j=\tau_j+d$, where $d=1$~s. Each excerpt starts with $e_{\mathrm{last}}=0$. If $e_{\mathrm{last}}$ is the end of the most recently accepted clip and $E_i$ the excerpt end, the acceptance indicator is
\begin{equation}
q_j=\begin{cases}
1,&a_j\ge e_{\mathrm{last}}\ \text{and}\ a_j+L\le E_i,\\
0,&\text{otherwise}.
\end{cases}
\label{eq:acceptance}
\end{equation}
An accepted request adds $[a_j,a_j+L)$ to $C_i$ and updates $e_{\mathrm{last}}$; a rejected request is not queued. Each run supplies one candidate, and invalid samples reset the run. Thus smoothing changes the number and timing of eligible runs without changing clip length or filling missing data.

Afterwards the experimenter's markers partition the requested footage into task, labelled rest, and unlabelled context. For block $i$ with clip union $C_i$ and task interval $T_i$, requested time $R$, retained task time $H$, and task coverage $\rho$ are
\begin{equation}
\begin{aligned}
R&=\sum_i|C_i|, & H&=\sum_i|C_i\cap T_i|,\\
\rho&=H\big/\sum_i|T_i|.
\end{aligned}
\label{eq:metrics}
\end{equation}
We also report $B_q$, the number of blocks keeping at least $q$ seconds of task. A whole-task interval is counted as recovered when a positive run overlaps the task interval with intersection-over-union of at least 0.5.

\subsection{Datasets and protocol}
\textbf{Sustained state.} Dataset~A of Shin \emph{et al.}~\cite{shin} contains 28-channel EEG with two EOG channels at 1000~Hz during an n-back working-memory task. We use participants VP002--VP006, three sessions each (VP001, used in development, is excluded). Each session has nine blocks of about 44~s separated by rest; the first six calibrate the model and the last three are held out, giving 45 evaluation blocks. Task onset and end come from the experimenters' own markers, which supply training and scoring labels, not classifier inputs; each held-out excerpt extends ten seconds before and after the task. This task-aligned origin carries timing information to the clock baselines, which also share the EEG-validity gate. The archived detector configurations use a circular time-shift null (2000 shifts per session), which preserves the autocorrelation that label shuffling would destroy. Models are frozen and hashed before any held-out block is scored.

\textbf{Electrodes and brief events.} OpenNeuro ds006394~\cite{leong,dataset} provides 16-channel OpenBCI EEG at 125~Hz from 33 people performing visual and auditory surprise tasks with the authors' own event labels. All 51 artifact-screened recordings enter the montage comparison (Table~\ref{tab:electrodes}). Per-recording five-fold cross-validation on 11 features per channel measures how well a wearer-calibrated model separates surprise from dummy-surprise epochs, with channel subsets emulating headbands and glasses-mounted electrodes. A leave-one-person-out variant tests an uncalibrated wearer; an archived sliding-window run tests transfer from stimulus-locked epochs to a continuous stream.

\textbf{Recorder.} An Android client polls a Python server every 250~ms and records H.264/AAC clips through Camera2, saving each clip locally before upload. End-to-end tests use synthetic EEG; one explicit-command cycle on a Samsung S23 produced a 10.02~s file.

\section{Experimentation and Results}
\label{sec:results}

\subsection{Does a simple sustained-state score carry signal?}
Across the fifteen held-out sessions the smoothed score reaches a mean ranking probability of 80\%, where the ranking probability is the chance that the score ranks a task window above a rest window and 50\% is chance level. In the separate archived detector configurations, participant means span 65--93\%; fourteen of fifteen sessions are above chance and 8 of 15 beat the circular-shift null at $p\le0.05$, with every participant represented in that set (Table~\ref{tab:capture}). The spread between people is real, two participants average 65\% and 67\%, and it is why the model is calibrated per wearer rather than using a population-fixed model. The archived eye-only control reaches 88\% in VP002 and 85\% in VP003, against 92\% and 80\% for EEG, so in two participants we cannot claim a purely cortical signal; we report both.

\subsection{What does smoothing do to the recorder?}
Smoothing helps the classifier: the ranking probability rises from 76.6\% to 79.8\%, and whole-task recovery doubles (10 to 21 of 45 blocks) because short positive runs merge. The recorder, however, measures footage (Table~\ref{tab:capture}). Merged runs mean fewer clips: 750~s instead of 980~s, a 23\% reduction overall, with reductions in every participant. Both versions keep at least five seconds of every task block. The cost is 902~s of task footage falling to 697~s. Fig.~\ref{fig:pipeline}(b) shows one such merge. A clock every 20~s keeps as much task as the raw score but asks for 14\% more film; a single clock at 25~s is cheap only because the task always starts ten seconds into the excerpt. Both clock baselines share the EEG-validity gate.

\begin{table}[t]
\caption{What the camera keeps on the same 45 held-out blocks}
\label{tab:capture}
\centering
\scriptsize
\setlength{\tabcolsep}{4pt}
\begin{tabular}{lccc}
\toprule
Policy & Ranking & Footage asked (s) & Task kept (s)\\
\midrule
Raw score & 76.6\% & 980 & 902\\
1~s smooth & \textbf{79.8\%} & \textbf{750} & 697\\
Clock every 20~s & -- & 1120 & 898\\
Clock once at 25~s & -- & 370 & 370\\
\bottomrule
\end{tabular}
\par\smallskip\raggedright\scriptsize Ranking probability: chance a task window scores above a rest window (50\% is chance level). Clocks have no EEG score. Available task time 1959~s.
\end{table}

\subsection{How many electrodes, and where?}
Table~\ref{tab:electrodes} compares scalp-layout proxies on all 51 surprise-task recordings, as a single ranking probability per electrode set. With a wearer-calibrated model, halving the montage from sixteen to eight electrodes in a headband-like layout costs two points (84\% to 82\%). Two electrodes at the midline (Fz, Cz) still reach 76\%, beating four electrodes at eyeglass positions (65\%). Of all 120 electrode pairs, every one of the twelve best contains Cz or Fz, and the best pair reachable from a glasses frame ranks 90th. A separate leave-one-person-out analysis of the parent cohort reaches 71\% with the full montage and 60\% with glasses pairs; its training protocol and cohort differ from this comparison. These scalp-layout proxies favor headband sensing over glasses-reachable contacts. This comparison uses the surprise task, not the n-back task of the recorder evaluation, so it does not by itself fix the workload montage.

\begin{table}[t]
\caption{Ranking performance by electrode count and placement}
\label{tab:electrodes}
\centering
\footnotesize
\setlength{\tabcolsep}{3.5pt}
\begin{tabular}{clc}
\toprule
Electrodes & Placement & Ranking\\
\midrule
16 & Full scalp (reference) & \textbf{84\%}\\
8 & Headband layout$^{\dagger}$ & \textbf{82\%}\\
4 & Fz Cz F7 F8 & 76\%\\
2 & Midline: Fz Cz & 76\%\\
2 & Frontal: F3 F4 & 73\%\\
4 & Glasses: Fp1 Fp2 F7 F8 & 65\%\\
2 & Glasses brow: Fp1 Fp2 & 63\%\\
2 & Glasses temples: F7 F8 & 61\%\\
\bottomrule
\end{tabular}
\par\smallskip\raggedright\scriptsize Ranking probability: chance the wearer-calibrated model ranks a surprise epoch above a control epoch (five-fold cross-validation within each recording). $^{\dagger}$F3 F4 C3 C4 P3 P4 O1 O2, a headband proxy drawn from the recorded channels, not a measurement of that device.
\end{table}

\subsection{From epochs to streams}
The surprise model separates stimulus-locked epochs, but a camera has no stimulus clock. Run as a sliding 0.8~s window over the continuous recordings, with its threshold and cooldown chosen on other people, the same classifier detected 1.5\% of events at 29 false triggers per hour. Part of this is implementation (epoch baseline correction and streaming timestamps differ). Discrimination at a known onset did not transfer to discovering onsets in this test. Training on the visual task and testing on the same person's auditory task likewise dropped the ranking probability from 78\% to 63\%. The evaluated trigger therefore reacts to a sustained state, calibrated on the activity the wearer will actually perform.

\subsection{What the EEG tells us}
Four points survive the experiments and guide the recording design. First, sustained task/rest discrimination is detectable in the recorded EEG with a linear band-power model, and detectability varies enough between people to motivate per-wearer calibration. Second, the tested instantaneous-event detectors did not establish a reliable trigger; the unit of decision is a state that lasts seconds. Third, smoothing is a recording-policy choice, not a free accuracy gain: it quietens the trigger and keeps a minimum of context in every block, but it also shortens what is kept, a trade best stated to the wearer in seconds of footage. Fourth, on the surprise task, electrode placement dominates electrode count: eight electrodes spanning the scalp outperform two frontal contacts at glasses-reachable positions. Whether the same ranking holds for sustained workload is untested.

\subsection{Limitations}
The offline model distinguishes task from rest in the evaluated recordings. Whether a captured interval is meaningful or useful to the wearer remains untested. A live evaluation also needs causal preprocessing and synchronised EEG and camera timestamps.

\section{Conclusion}
\label{sec:conclusion}
In offline replay, a sustained task/rest score estimated with a band-power model can drive recording requests. A one-second smoothing half-life improved score ranking while reducing both total recording demand and retained task time. Every block still contributed at least five seconds of task footage. These results show why the effects of smoothing should be measured at the recording decision as well as at the classifier. In the separate surprise task, eight well-placed electrodes reached 82\% ranking performance against 84\% for the full montage, motivating an electrode-placement study for workload. The tested brief-event detectors did not significantly outperform the random-placement null. The next steps are to repeat the montage comparison on workload EEG and test a causal pipeline with synchronised EEG, ocular controls, and camera acknowledgements, then ask wearers whether the clips were worth keeping.

\end{document}